# 2 inch Molecular Organic Glass Scintillator for Neutron-Gamma Discrimination

**M. Grodzicka- Kobylka** [a] [1], **T. Szczesniak** [a], **M. Moszyński** [a], **L. Swiderski** [a], **K. Brylew** [a], **P. L. Feng** [b], **L. Nguyen** [b], **J.S. Carlson** [c], **J.J. Valiente-Dobón** [d], **J. Trzuskowski** [a], **A. Misiarz** [a], **Ł. Talarek** [a], **P. Zając** [a]

[a] *National Centre for Nuclear Research, A. Soltana 7, PL 05-400 Świerk-Otwock, Poland*

[b] *Sandia National Laboratories, 7011 East Ave. Livermore, CA 94550, United States of America*

[c] *Blueshift Optics LLC, 2744 E 11th St., Ste H2, Oakland, CA 94601-1443, United States of America*

[d] *INFN, Laboratori Nazionali di Legnaro, I-35020 Legnaro, Italy*

*E-mail*: `Martyna.Grodzicka@ncbj.gov.pl`

ABSTRACT: In this manuscript we report on the scintillation properties and pulse shape discrimination (PSD) performance of new organic glass scintillator. Two cylindrical samples with dimensions of 2×2 inches were tested. Additionally, this two samples were used in stack configuration in order to measure the PSD characteristics of a sample with a size of 2 × 4 inches. The study covers the measurements of neutron/gamma discrimination capability, emission spectra, photoelectron yield and analysis of the light pulse shapes originating from events related to gamma-rays and fast neutrons. The results were compared to data recorded previously using an EJ-276 plastic scintillator, an EJ-309 liquid scintillator and a stilbene single crystal.

---

[1] Corresponding author.

# 1 Introduction

The current development of scintillator technology allows to meet more and more stringent requirements for fast neutron detection. Currently, the most important features of fast neutron detectors include:

- good discrimination between detected neutrons and gamma-rays with a low energy threshold;
- non-limited size of detector with unaffected performance;
- high detection efficiency for fast neutrons;
- resistance to high radiation doses;
- durability (stability of response over time, low temperature dependence);
- safe maintenance (low toxicity, inflammability)
- low material cost

The possibility of producing large-size scintillators (low self-absorption effect in the scintillator). Good n/γ discrimination can be achieved due to the efficient conversion of absorbed radiation into visible light (light yield) and the different proportions of prompt and delayed components in scintillation pulses produced by both neutron and gamma interactions. The pulses induced by both types of radiation contain both a short decay (immediate) and long (delayed) fluorescent components [1, 2]. Currently, the best low energy threshold differentiation between neutrons and gamma rays can be obtained using single stilbene crystals. However, the largest stilbene single crystals reach size of 5 × 3 inches or 4 × 4 inches (offered by Inrad Optics [3]) and the scintillators become increasingly fragile as the dimensions increase, making it unsuitable for industrial use. The second group of the most popular detectors with still good neutron and gamma ray discrimination are liquid scintillators, with the most popular being NE213, originally produced by Nuclear Enterprise, or its equivalents BC501A, offered by Saint-Gobain Crystals and EJ-301, commercially available from Eljen Technologies. Unfortunately, these xylene based detectors are known for their toxicity and flammability, since they have a low flash point of about 25 °C. To overcome these drawbacks, EJ-309 scintillator has been developed as an alternative to NE213. EJ-309 has a high flash point of about 144 °C and a low vapor pressure of about 0.002 mm Hg. at 20 °C [4]. Furthermore, EJ-309 has low chemical toxicity and is compatible with cast acrylic plastics, which makes it suitable for use in difficult environmental conditions.

Since exploiting liquid scintillators has still been considered unsafe in some applications due to the probability of uncontrolled leakage efforts were made towards inventing plastic scintillators with n/γ discrimination capability (henceforth called "n/γ plastics"). The first commercially available n/γ plastics were developed by Zaitseva et al. from Lawrence Livermore National Laboratory and are currently offered as EJ-276 by Eljen Technology [5]. However, the results of n/γ discrimination obtained with plastic detectors are significantly worse than those obtained with stilbene crystals or liquid scintillators.

Since 2016, a new scintillator with the ability to differentiate gamma neutrons has appeared on the market, namely small-molecule organic glass scintillator (hereafter OGS). This scintillator was developed by Carlson & Feng from Sandia National Laboratories [6].

The aim of this research was to study the properties of the latest developed OGS sample. This paper reports the results of measurements on n/γ discrimination and the photoelectron yield, light output, emission spectra and analysis of the light pulse shapes originating from events related to gamma-rays and fast neutrons. Additionally, two samples of the 2 × 2 inch OGS scintillators were tested in a stack to check the PSD capability of larger sample 2 × 4 inches. The results were compared to a latest available PSD plastic EJ-276 [7], a liquid scintillator EJ-309 and a stilbene single crystal. The errors of the results presented here were determined by calculating the standard deviation of the series of measurements performed in the same conditions.

## 2 Experimental details

### 2.1 Scintillators and photomultipliers

The OGS scintillator, plastic scintillator EJ-276 and Stilbene single crystal used in the tests were polished on all surfaces and wrapped with Teflon tape on all sides, except for those that were kept transparent in order to transfer light to the PMT photocathode. The EJ-309 liquid scintillator had a typical form of aluminum cylindrical containers. The inner sides of the cells were lined with white reflector. The containers were sealed with a glass window on one of the sides. The list of the tested scintillators is presented in Table I.

The bare optical surfaces of the tested scintillators were coupled to a commercially available spectrometry PMT Hamamatsu R6233-100 using silicone grease (Baysilone Ol M 600 000 cSt). The R6233-100 PMT has a 76 mm diameter photocathode and is characterized by high photocathode blue sensitivity of 15.4 µA/lmF that results in high quantum efficiency of 41% according to the datasheet provided by with the unit used [8].

Four different radioactive sources were used in this work to record gamma ray and neutron response of the tested samples: $^{241}$Am (59.5 keV), $^{137}$Cs (661.7 keV), $^{22}$Na (511, 1274.5 keV) and PuBe (4.4 MeV).

TABLE I
MAIN PROPERTIES OF SCINTILLATORS USED DURING THE STUDIES

| Crystal | diameter x height (inch) | Shape | Peak emission [nm] | Type of scintillator | Manufacturer |
|---|---|---|---|---|---|
| OGS | 2x2 | Cylinder | 428 | glass | Sandia National Laboratories |
| OGS | 2x4 | Cylinder | 428 | glass | Sandia National Laboratories |
| EJ-276 | 2x2 | Cylinder | 425 | plastic | Eljen Technology |
| EJ-309 | 2x2 | Cylinder | 424 | liquid | Eljen Technology |
| Stilbene | 3x1 | Cylinder | 430 | single crystal | Inrad optics |
| Stilbene | 3x3 | Cylinder | 430 | single crystal | Inrad optics |

### 2.2 Emission spectra

Although the peak emission of organic scintillators as tested in this study is provided in the datasheets, we decided to measure the emission spectra of the samples in order to precisely determine their light output. The emission spectra measurements were performed by focusing the light from the samples on the entry slit of a PC-controlled Digikröm CM110 monochromator. A 1 GBq $^{241}$Am source was used to irradiate the samples with 59.5 keV γ-rays. On the other side of the monochromator a side-on Hamamatsu R928P PMT with calibrated quantum efficiency (QE) was placed, operated in the single photon counting mode. Light from the tested scintillators was diffracted by the grating system of the monochromator and collected by the PMT. The PMT anode signal was fed to a discriminator, and then sent to an Ortec 994 multiscaler where single photoelectron pulses were registered. The multiscaler was synchronized with the monochromator using laboratory-grade software. The accuracy of the monochromator was 8 nm with 1.0 mm slits and 1 200 grooves per millimeter grating. A diagram of the experimental setup is presented in Fig. 1.

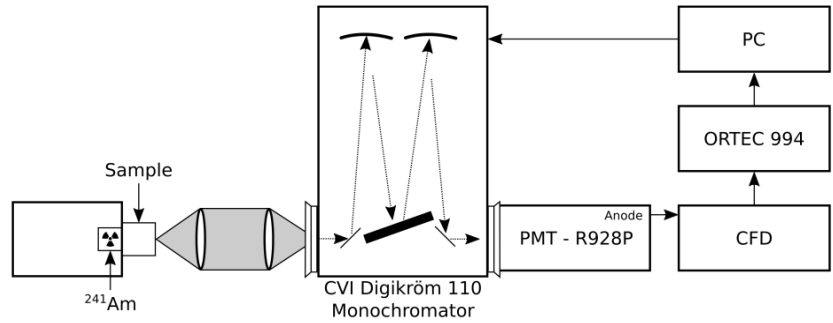

**Fig. 1.** The diagram of the experimental setup used to record emission spectra.

## 2.3 Photoelectron yield

The number of photoelectrons was estimated using the Bertolaccini et al. method [9]. This method is based on a comparison of the scintillation peak centroid due to a given gamma-ray energy with the peak centroid of a single photoelectron. All tested scintillators do not show full energy peaks above a certain energy threshold, as they are not capable of absorbing the entire γ-ray energy. This is due to the low density and low atomic number of light material based scintillators. Thus the energy calibration is done by registering a Compton continuum of gamma-rays (from $^{137}$Cs source tests reported herein), and then evaluating a Compton edge position (assumed to be at a channel where spectrum height reaches 80% of the upper part of the Compton distribution [10, 11]) and a single photoelectron spectrum from a PMT.

Light output is the ratio between the number of recorded photoelectrons and integral quantum efficiency (IQE), where the IQE is an average of QE weighted by the emission spectrum function [12].

## 2.4 Pulse shape discrimination

Discrimination of neutrons from γ-rays is often done due to differences in the pulse shapes induced in the detectors by these forms of radiation, thus it is usually called "pulse shape discrimination" (PSD). In the course of measurements reported herein we used the PMT output signal that drove a DT 5730 waveform digitizer, which has a 500 MHz sampling rate (2 ns per sample), a 14-bit resolution and a 2 V dynamic range. For PSD analysis the PuBe source was used. A schematic of experimental setup is presented in Fig. 2.

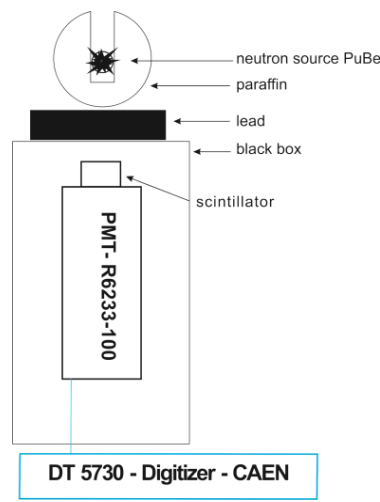

**Fig. 2.** Experimental setup for the PSD measurements.

In this work was used the Charge Comparison Method which is a classic PSD method, based on a comparison of two different integrals of the current signal. PSD parameter was defined as a ratio of difference of integrals of a long gate and a short gate divided by the integral of a long gate:
($Q_{long\ gate}$-$Q_{short\ gate}$)/$Q_{long\ gate}$), see Fig. 3. The formula was implemented in a CAEN DT5730 digitizer [13].

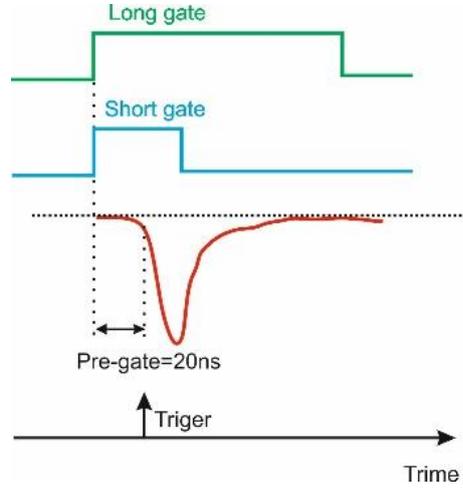

Fig. 3. Diagram of the n/γ discrimination method implemented in a CAEN DT5730 digitizer. The pre-gate of 20 ns contributes to the length of each gate.

In order to quantify the PSD performance of the detectors, the figure of merit (FOM) was calculated as follows:

$$FOM = \frac{(peak\ separation)}{(FWHM_\gamma)+(FWHM_n)} \quad (1)$$

where FWHM stands for the full width at half maximum of the peaks corresponding to neutron and γ-ray detection as projected onto the PSD parameter axis. FOM for each pulse was evaluated for narrow energy ranges (E±10%E) set at: 100 keVee, 300 keVee, 500 keVee and 1000 keVee. Additionally, FOM was calculated also for a wide energy range between 100 keVee and 1000 keVee. $^{241}$Am, $^{137}$Cs and $^{22}$Na γ-ray emitters were used for the energy calibration of 2D plots, providing high accuracy for selection of energy window cuts.

## 2.5   Light pulse shapes from gamma-rays and fast neutrons

A detailed diagram of the slow-fast experimental setup for capturing scintillation decay profiles is presented in Fig. 4. The light pulse shapes were recorded using the Bollinger-Thomas single photon method [14] modified to address the study of a light pulse shape due to gamma-rays and fast neutrons separately [15]. Two photomultipliers, i.e. a Photonis XP20D0 (coupled to the tested OGS scintillator) and a Hamamatsu R5320, were used to detect single photons emitted from the tested sample. The PMTs were placed on opposite sides of a light-tight tube. Only the sides of the tested sample were covered by Teflon tape, whereas the top and the bottom of the sample were open, thus allowing the detection of scintillation photons by both PMTs. The low time jitter of the Hamamatsu R5320 timing PMT, equal to 140 ps [16], enabled sufficient time resolution for precise registration of the scintillation decay profiles, particularly its fast component. A detailed description of the method used to record light pulses is presented in [5].

The scintillators were irradiated by a PuBe neutron source. A 10 cm lead brick was placed between the scintillator and the PuBe source in order to suppress gamma-rays from the neutron source, which could produce Cherenkov photons in the PMT window. The recorded time distribution of single photons detected by the R5320 PMT reflects the scintillation light pulse shape.

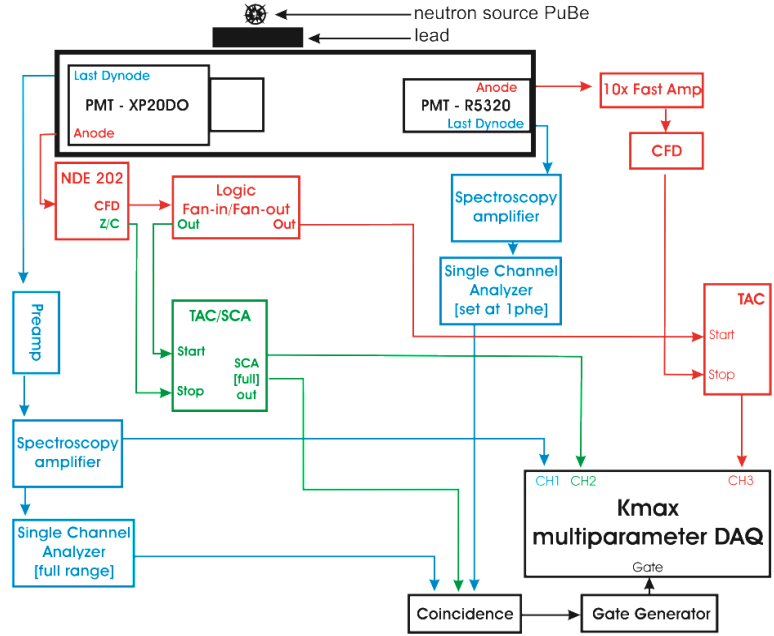

**Fig. 4.** Experimental setup for the light shape measurements using the Bollinger-Thomas single photon method with PSD, following Ref. [15].

## 3 Results

### 3.1 Emission spectra

The emission spectra of scintillators OGS, EJ-276, EJ-309 and stilbene are presented in Fig. 5. The data are collected in Table II. All of the spectra were corrected for the quantum efficiency of the R928P PMT and the monochromator's grating.

It is worth noting that the emission spectrum of the OGS scintillator is similar to the emission spectra of the EJ-276 plastic and EJ-309 liquid scintillator. Their emission spectra have two main components, one peaked at ca. 430 nm, while the second at about 460 nm. The main difference among the spectra is the intensity of the second peak - the highest intensity was observed for the EJ-276 plastic scintillator, whereas the lowest intensity was observed for the OGS scintillator. The maximum of the emission spectrum for stilbene crystal is shifted in relation to the other reported here scintillators and has a maximum at 383±15 nm.

In order to quantify the properties of the tested samples in terms of light output, integral quantum efficiency (IQE) was calculated for all of the tested detectors as coupled with the calibrated R6233-100 PMT (see Table II). The Quantum Efficiency for R6233-100 PMT is presented in Fig. 6. The IQE is a measure of a scintillator-photodetector spectral matching, allowing to convert recorded number of photoelectrons into photons collected from a sample. Stilbene crystal has the highest value of IQE equal to about 39%, for the other scintillators this value ranged between 30 and 33%. According to Fig. 6, in the 340-410 nm range, the QE of the photomultiplier used has the highest value of around 40%, in this range there is also the maximum of the emission peak of the stilbene scintillator. The additional structure at 460±16 nm occurring in a scintillators OGS, EJ-276 and EJ-309 is already on the slope of the QE curves, which results in a worse match.

Since the differences in the IQE values between the OGS, EJ-276 and EJ-309 detectors are small, it can be assumed that the photodetector is equally well suited to these tested scintillators, and differences in n/γ discrimination results come only from the properties of scintillators manifested in the decay profiles. In the case of the stilbene crystal, the PMT's QE is much better suited to the detector, which makes n/γ discrimination also privileged in its case.

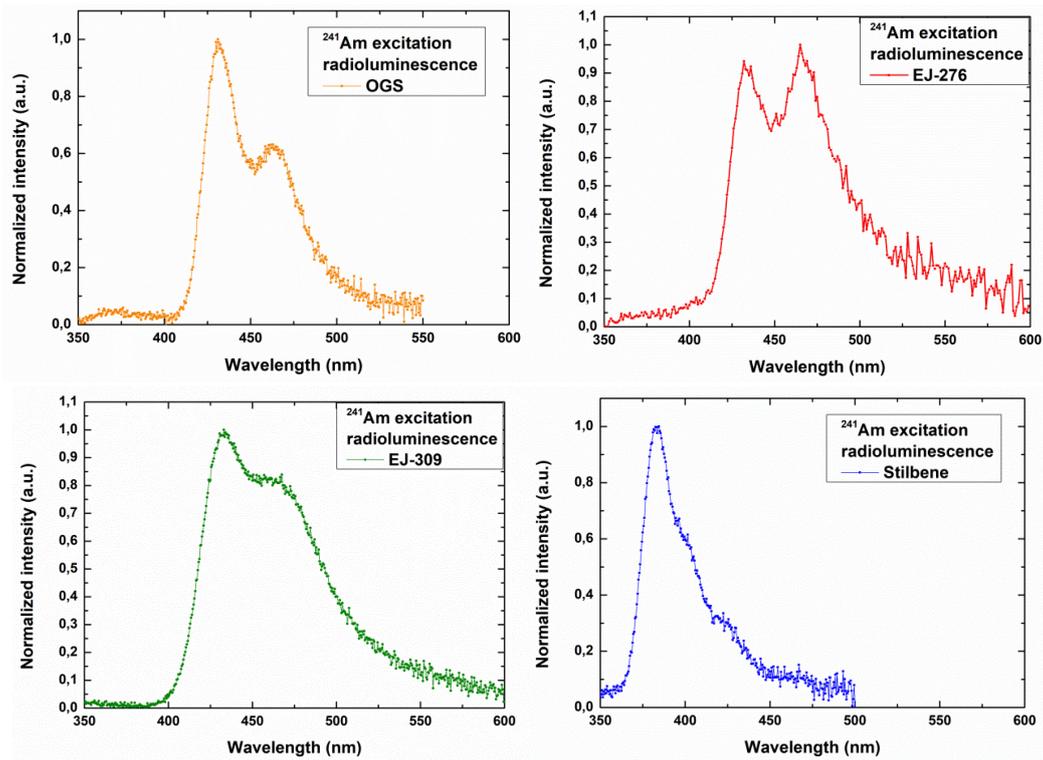

**Fig. 5.** Emission spectra of the OGS, EJ-276, EJ-309, and Stilbene recorded with the calibrated R928P PMT. Results were corrected for quantum efficiency of the PMT and for the spectral response of the grating system.

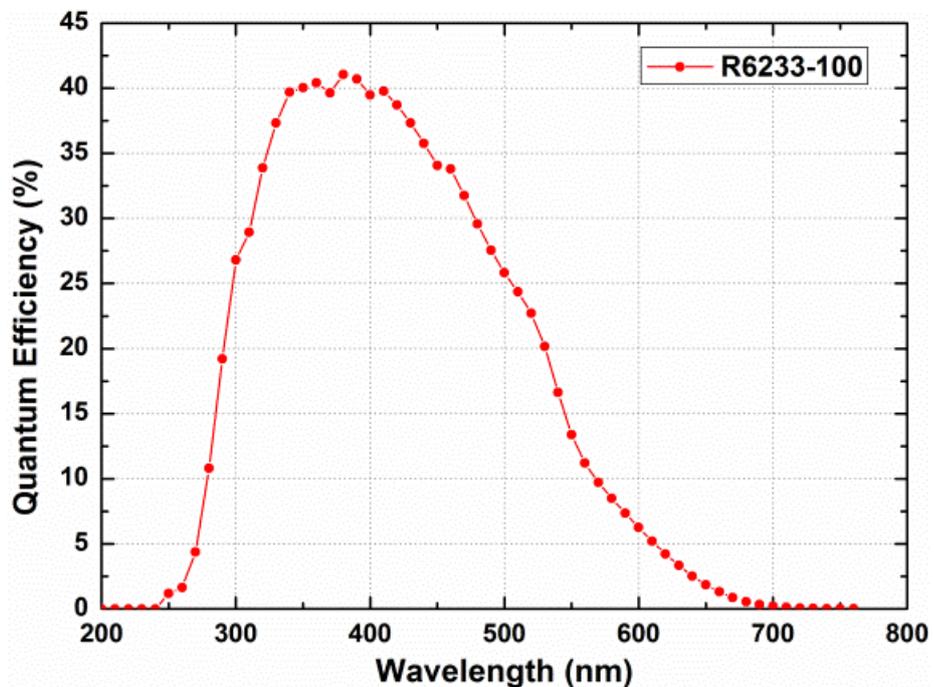

**Fig. 6.** Quantum Efficiency of the R6233-100 PMT used in this study.

**Table II**
Scintillation properties of the tested scintillators

| Scintillator | Wavelength of Maximum Emission (nm) | | Integral QE (R6233-100+emission spectrum of scintillator) |
|---|---|---|---|
| | First maximum | Second maximum | |
| OGS | 431±16 | 463±16 | 33 ± 3% |
| EJ-276 | 434±16 | 466±16 | 30 ± 3% |
| EJ-309 | 433±16 | 460±16 | 31 ± 3% |
| Stilbene | 383±15 | - | 39 ± 3% |

### 3.2 Photoelectron yield

The number of photoelectrons per MeV and light output (LO) for all of tested samples are presented in Table III. OGS scintillator has the highest number of photoelectrons per MeV and the highest LO of all tested scintillators. It emits ca. 25% more light than the 3×1inch Stilbene single crystal.

Up until now, the most crucial limitation for the application of scintillators in neutron/gamma discrimination has been the deterioration of their PSD parameters as their size increased, mostly due to the self-absorption of light. In order to shed some light on this issue, the characterization of PSD performance with respect to the scintillator height was done using two samples of the OGS scintillator, coupled in a stack using silicone grease and then wrapped with several layers of Teflon tape (0.2 mm thickness and 0.5g/cm$^3$ density). The studied samples had a diameter of 2 inches and height of from 2 to 4 inches. In these configurations, even without self-absorption, the expected number of photoelectrons per MeV should be slightly reduced compared to that of a single scintillator due to light loss at the interfaces between the individual samples, affected also by the quality of the surface polishing. This approach allowed us to investigate the self-absorption problem. The results obtained are reported in Table III. The light output decreased by about 35% for the 2×4 inch sample compared to the 2×2 inch sample. It is worth noting that the EJ-276 scintillator lost more than 20% of light in the 3 years following its production.

**Table III**
Photoelectron yield and light output of investigated detectors.

| Scintillators | diameter x height (inch) | Phe number (phe/MeV) | LO (Ph/MeV) |
|---|---|---|---|
| OGS (sample 1 sample 2) | 2"x2" | 6730 ± 200<br>6960 ± 200 | 20 200 ± 2 000<br>20 900 ± 2 000 |
| OGS | 2"x4" | 4450 ± 140 | 13 400 ± 2 000 |
| EJ-276 (sample from 2017) | 2"x2" | 2450 ± 70<br>(in 2017- 3140 ± 150) | 8 200 ± 1 000<br>(in 2017- 10 500±1000) |
| EJ-309 | 2"x2" | 4100 ± 130 | 13 200± 1 100 |
| Stilbene | 3"x1" | 6400 ± 190 | 16 500± 1 600 |
| | 3"x3" | 4420 ± 140 | 11 400± 1 200 |

### 3.3 Pulse shapes

The Thomas-Bollinger method was used to record scintillation pulse shapes in a mixed field of fast neutrons and γ-rays. In order to record a separated response of the tested scintillators to the two types of radiation, PSD and energy data were used to tag the events giving rise to specific scintillation pulses. The light pulses recorded using OGS detector, after applying the constraints on energy and PSD parameter, are presented in Fig. 7.

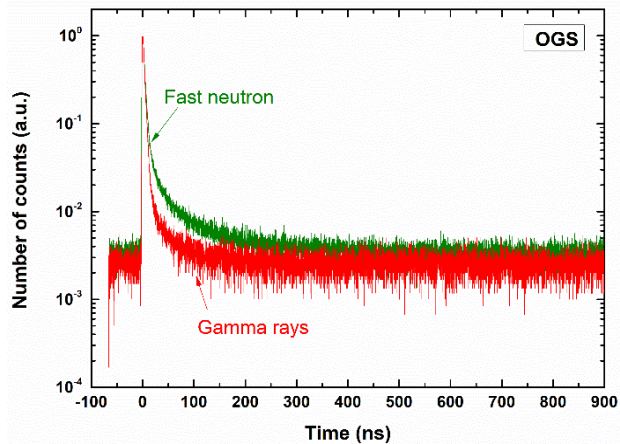

**Fig. 7.** Normalized light pulse shapes obtained for fast neutrons and γ-rays recorded with OGS detector. The TAC range was set to 1 000 ns.

As can be observed, OGS detectors show different decay profiles for γ-rays and fast neutrons. This feature allows to use the PSD for discriminating fast neutrons from γ-rays. In order to compare the OGS decay times of the scintillator with earlier measurements of EJ-276 and EJ-309 scintillators made under the same conditions, the components the light pulses were fitted with triple exponential curves in OriginPro 8.6 software [17]:

$$y = y_0 + A_1 exp(-x/\tau_1) + A_2 exp(-x/\tau_2) + A_3 exp(-x/\tau_3) \qquad (2)$$

where: $A_1$, $A_2$ and $A_3$ are the amplitudes of the exponential functions, $\tau_1$, $\tau_2$ and $\tau_3$ are the scintillation decay time constants, and $y_0$ is the baseline offset originating from random coincidences. The contribution of random coincidences was determined from the time range in the TAC spectrum preceding the light pulse leading edge. The three exponential components observed in the organic glass scintillator are related to fast, medium and long components. In the case of previously characterized detectors EJ-309 and EJ-276 we had to use four exponential components in order to fit the scintillation decay profiles. The results of the fits are presented in Table IV. The light pulse shapes measured for γ-rays and fast neutrons, together with the multi-exponential fits calculated for the tested OGS scintillator are presented in Fig. 8. Each plot contains experimental data extracted from the measured spectra after selection of coincidence windows that correspond to events induced by either neutrons or gamma-rays. Individual exponential components fitted to each scintillation decay profile are also presented. Scintillation decay data for EJ-309, EJ-276 and stilbene single crystal were obtained using the same experimental method and nearly identical setup as reported in [15] and [7], respectively.

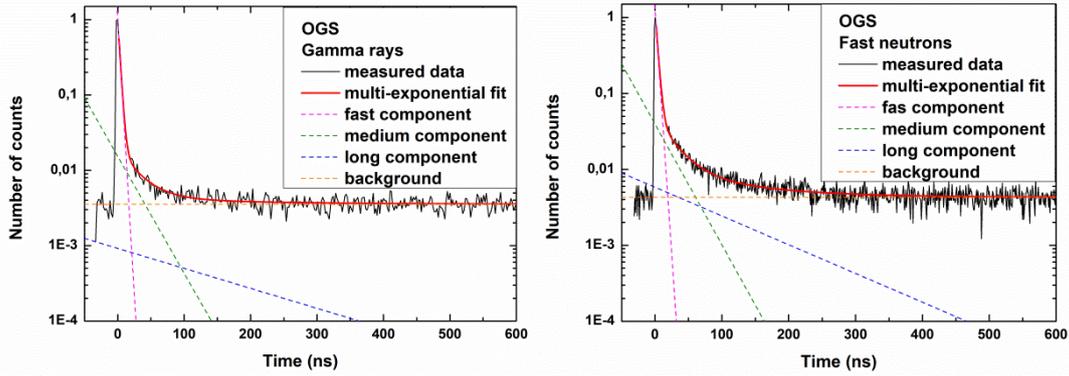

**Fig. 8.** Light pulse shapes recorded for the tested OGS (gamma rays- at the left, fast neutrons- at the right). Experimental data are presented with solid lines, whereas fitted decay components are shown with dashed lines.

**Table IV**
Fitting parameters of all components calculated for the tested plastic scintillators

| scintillator | radiation | Fast component | | Medium component | | Long component 1 | | Long component 2 | |
|---|---|---|---|---|---|---|---|---|---|
| | | Decay const. [ns] | Intensity % | Decay const. [ns] | Intensity % | Decay const. [ns] | Intensity % | Decay const. [ns] | Intensity % |
| EJ-309[a] | Gamma | 3.7±0.4 | 80 | 31±3 | 10 | 140±10 | 7 | 790±80 | 3 |
| | Fast N | 4.8±0.5 | 46 | 32±3 | 24 | 140±10 | 20 | 620±60 | 11 |
| EJ-276[b] | Gamma | 4.0±0.4 | 70 | 16±2 | 12 | 98±10 | 8 | 690±70 | 8 |
| | Fast N | 3.9±0.4 | 47 | 18±2 | 13 | 106±10 | 13 | 800±80 | 27 |
| Stilbene | Gamma | 5.5±0.5 | 86 | 49 ±5 | 9 | 330±30 | 6 | - | - |
| | Fast N | 6.6±0.5 | 52 | 47 ±5 | 21 | 260±30 | 26 | - | - |
| OGS | Gamma | 3.3±0.3 | 89 | 29 ±3 | 8 | 158±20 | 3 | - | - |
| | Fast N | 3.7±0.3 | 70 | 27 ±3 | 15 | 100±10 | 16 | - | - |

[a] from [15], [b] from [7],

The n/γ discrimination capability of scintillators is described by the ratio of the intensities of the slow components due to neutrons (proton recoil), to gamma rays (primary electrons). The data in Table IV allows for the calculation of the respective ratios for all tested scintillators (OGS, Stilbene single crystal, EJ-276 and EJ-309). Taking into account the sum of the medium and the long components in the 16 ns to 800 ns ranges, results in the values of 2.8, 1.9, 3.1 and 2.8 for EJ-309, EJ-276, Stilbene and OGS were obtained respectively. Thus confirming stilbene superiority than the others tested scintillator, and placing the OGS on par with the EJ-309 scintillator.

### 3.4 PSD performance

In this part of the study we compare n/γ discrimination performance of OGS scintillator with EJ-276 plastic, EJ-309 liquid and stilbene scintillator. Each of the tested scintillators was characterized after individual optimization procedure. The optimization consisted in determining the length of the short and long gate, for which the obtained FOM value for the tested scintillator was the highest. The length of the short gate was changed in 10 ns steps from 50 ns to 90 ns, while the length of the long gate was changed in 100 ns steps from 200 ns to 1200 ns. The pre-gate was set to 20 ns (see Fig. 3). Optimization results for the OGS scintillator are shown in Fig. 9. The best value of FOM was achieved for 70 ns short gate and 400 ns long gate. FOM for 100, 300, 500 and 1000 keVee energy cuts are 1.41, 2.59, 3.0 and 3.4 respectively.

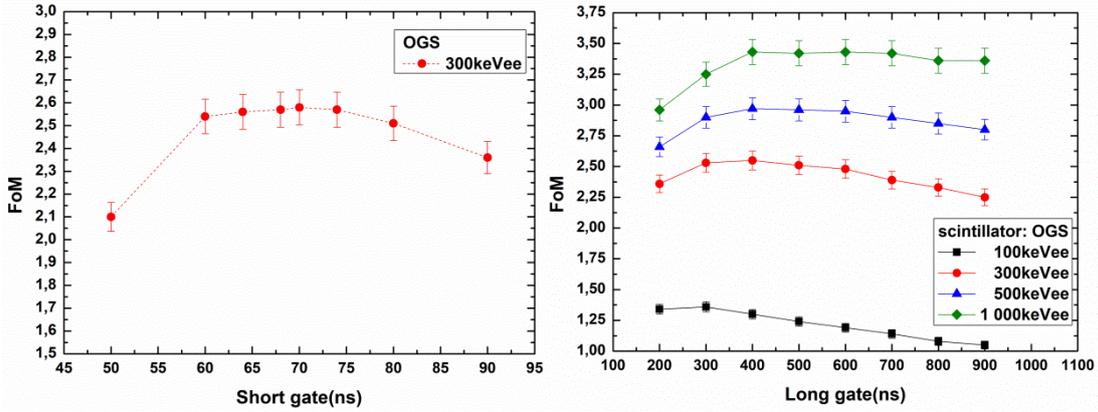

**Fig. 9.** The FoM calculation as a dependence of short integration gate and long integration gate for different Energy window.

The PSD capabilities of OGS samples were compared to the n/γ discrimination performance of the EJ-276, EJ-309 and stilbene scintillators (see Fig.10). All tested scintillators present a very clear separation of fast neutron and gamma-ray events down to 100 keVee. The numerical FOM results measured at energy cuts set at 100, 300, 500, and 1000 keVee and energy window between 100-1000 keVee of recoil electron energies show superior performance of the stilbene single crystal (see Table V). However, it is worth noting that the applied PSD method results in the OGS scintillator performance being almost as good as of EJ-309 liquid scintillator, making OGS a useful alternative for fast neutron detection, especially in demanding environmental conditions, where solid state detectors are preferred over liquid or gaseous detectors.

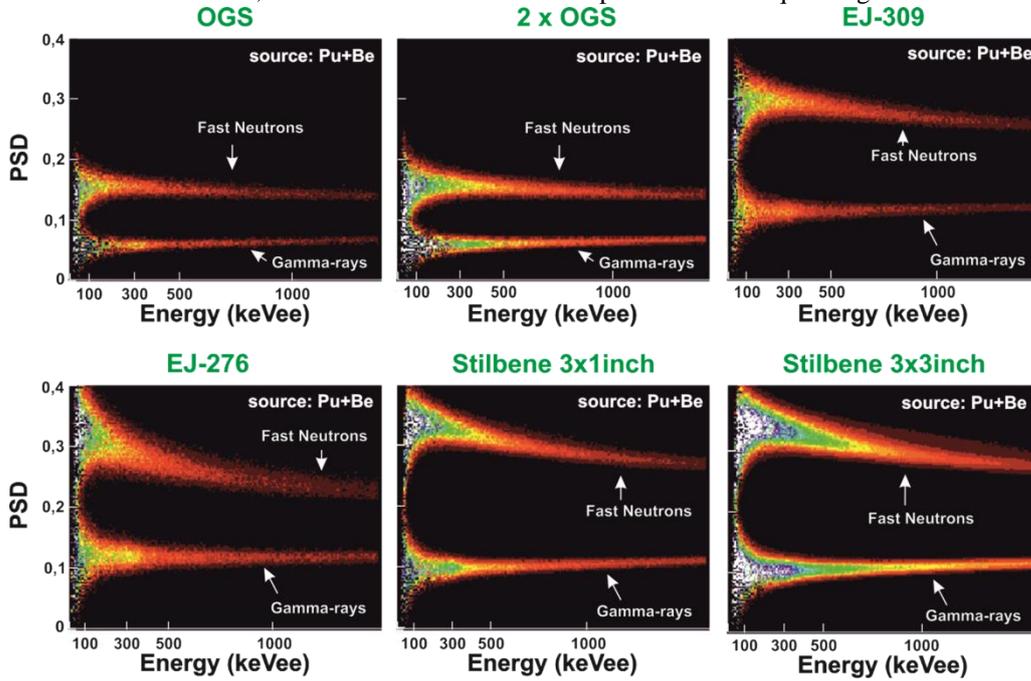

**Fig.10.** 2D plots of PSD versus energy recorded with PuBe sources for OGS, EJ-309, EJ-276 and Stilbene.

**Table V**

FOM for energy cuts at 100, 300, 500 and 1000 keVee and an energy window between 100-1000 keVee measured with EJ-309, EJ-276, stilbene single crystal and OGS scintillators.

| Scintillator | OGS (2x2inch) | 2x OGS (2x4inch) | EJ-309 (2x2inch) | EJ-276 (2x2inch) | Stilbene (3x1inch) | Stilbene (3x3inch) |
|---|---|---|---|---|---|---|
| Short gate Long gate | 70 400 | 70 400 | 60 800 | 74 700 | 70 800 | 70 800 |
| 100 keVee | 1.41±0.04 | 1.22±0.03 | 1.48±0.04 | 1.27±0.03 | 2.07±0.06 | 1.82±0.05 |
| 300 keVee | 2.59±0.07 | 2.18±0.06 | 2.88±0.08 | 2.06±0.06 | 3.8±0.1 | 3.4±0.1 |
| 500 keVee | 3.0±0.1 | 2.57±0.07 | 3.3±0.1 | 2.16±0.06 | 4.5±0.1 | 3.8±0.1 |
| 1000 keVee | 3.4±0.1 | 2.81±0.08 | 3.7±0.1 | 2.33±0.07 | 5.1±0.1 | 4.4±0.1 |
| 100-1000keVee | 2.17±0.06 | 1.93±0.05 | 2.34±0.07 | 1.32±0.04 | 2.81±0.06 | 2.68±0.08 |

Table VI summarizes the FOM for an energy window of 300 keV, normalized to the number of photoelectrons, according to publications [18] and [19]. The normalized FOM values are similar for the same type of scintillator, regardless of its size, which is consistent with previous publications and confirms the correctness of the measurements made.

**Table VI**

A comparison of PSD capabilities of the all tested scintillators at 300 keV energy cut.

| Scintillator | OGS | 2x OGS | EJ-309 | EJ-276 | Stilbene 3x1inch | Stilbene 3x3inch |
|---|---|---|---|---|---|---|
| Nphe at 300 keV | 2020±100 | 1340±70 | 1230±60 | 740±40 | 1920±100 | 1330±70 |
| FOM at 300 keVee | 2.59±0.07 | 2.18±0.06 | 2.88±0.08 | 2.06±0.06 | 3.8±0.1 | 3.4±0.1 |
| FOM/$\sqrt{N_{phe}}$ | 0.058±0.004 | 0.060±0.004 | 0.082±0.005 | 0.076±0.005 | 0.087±0.006 | 0.093±0.006 |

In Figure 11, we can see the FoM obtained for the OGS scintillator by other researchers. These results cannot be directly compared with each other because the PSD properties depend not only on the type of scintillator used, but also on its shape (diameter, size) and on the photodetector used to read the light obtained. A comparison of the PMTs employed to collect these data is presented in Table VII. In our case, we used the spectroscopic photomultiplier R6233-100 with a very high QE equal to 41%, the remaining researchers focused on timing photomultipliers with much lower QE, which in the case of publications [6, 20] explains the worse measured FoM values. Higher FoM obtained in ref. [6], despite the use of a photomultiplier with a lower quantum efficiency (QE=26%), are the result of using a smaller scintillator. Most likely, we observe a stronger self-absorption effect for larger size scintillator samples.

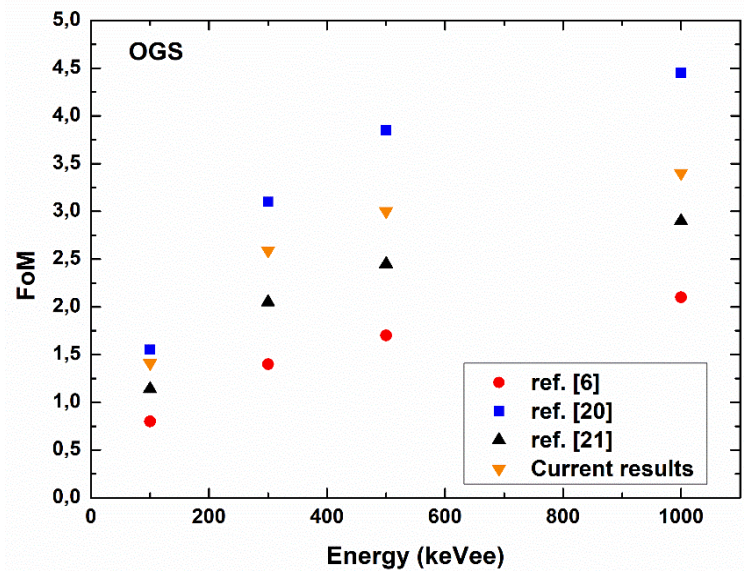

**Fig.11.** Figure of merit (FOM) for OGS scintillator.

**Table VII**
A comparison of used PMT

| Paper | Scintillator diameter x height (mm) | PMT | | | | | | | |
|---|---|---|---|---|---|---|---|---|---|
| | | Type | End window (mm) | Photocathode active diameter (mm) | Blue Sensitivity Index (CS 5-58) | QE (%) | Transit time (ns) | Rise time (ns) | Single electron jitter FWHM (ps) |
| Ref. [20] | 10x10 | ET 9124B | 29 | 25 | 11 | 26 | 33 | 3.0 | 5000 |
| Ref. [6] | 50.8 x 50.8 | ET 9214B | 51 | 46 | 11.5 | 30 | 45 | 2.0 | 2200 |
| Ref. [21] | 25.4 x 25.4 | ADIT L25D19 | 25 | 22 | 11.5 | 28 | 12 | 1.0 | 290 |
| Current article | 50.8 x 50.8 | R6233-100 | 76 | 70 | 15.4 | 41 | 52 | 6.0 | 8500 |

# 4  Conclusions

The new OGS scintillator developed by Sandia National Labolatory has many advantages. This includes an excellent light output about 20 000ph/MeV, an emission spectrum with maximum at about 430nm that is matched to most of the available photodetectors, fast decay of the pulse, which allows the measurement of larger neutron fluxes, and very good PSD. The highest speed and the intensity of the fast component of the OGS light pulse was compared to other detectors based on reference materials. This parameter is very important in fast timing in time-of-flight experiments. The PSD of OGS was measured to be better than for EJ-276 plastic scintillator and almost as good as for the EJ-309 liquid scintillator. However, the PSD performance obtained with a single Stilbene scintillator is still better. Taking into account the practical advantages (robustness, non-toxicity and non-flammability) of PSD organic glass scintillators over liquid scintillators, the OGS can be a reasonable alternative for neutron detection with gamma discrimination in many real-life applications, for example, in neutron/gamma ray handheld detectors.

### Acknowledgements
This work was supported in part by the European Union (ChETEC-INFRA, project no. 101008324).